\documentclass[12pt]{article}%
\usepackage{amssymb}
\usepackage{amsfonts}
\usepackage{amsmath}
\usepackage[nohead]{geometry}
\usepackage[singlespacing]{setspace}
\usepackage[bottom]{footmisc}
\usepackage{indentfirst}
\usepackage{endnotes}
\usepackage{graphicx}%
\usepackage{rotating}
\usepackage{parskip}
\makeatletter
\def\@biblabel#1{\hspace*{-\labelsep}}

\def\section{\@startsection {section}{1}{\z@}{-3.25ex plus -1ex minus 
    -.2ex}{1.5ex plus .2ex}{\large\bf\centering}}

\def\subsection{\@startsection{subsection}{2}{\z@}{-3.25ex plus -1ex minus 
   -.2ex}{1.5ex plus .2ex}{\large\it}}
\makeatother
\begin{document}

\title{\bf{The Doomsday Argument in Many Worlds}}
\author{\bf{Austin Gerig}\\
University of Oxford\\
austin.gerig@sbs.ox.ac.uk}

\date{ 
September 2012
\vspace{.1in}}

\maketitle
\vspace{-.15in}


You and I are highly unlikely to exist in a civilization that has produced only 70 billion people, yet we find ourselves in just such a civilization.  Our circumstance, which seems difficult to explain, is easily accounted for if (1) many other civilizations exist and if (2) nearly all of these civilizations (including our own) die out sooner than usually thought, i.e., before trillions of people are produced.  Because the combination of (1) and (2) make our situation likely and alternatives do not, we should drastically increase our belief that (1) and (2) are true.  These results follow immediately when considering a many worlds version of the ``Doomsday Argument'' and are immune to the main criticism of the original Doomsday Argument.


\thispagestyle{empty}

\newpage

\setcounter{page}{1}

\section{Introduction}

Imagine you are sitting at a table, blindfolded, and that an urn is placed in front of you.  You are told this urn can be one of two types: it is either a \emph{small urn} or a \emph{large urn}.  If it is a small urn, it contains 10 balls numbered 1 through 10.  If it is a large urn, it contains 1 million balls numbered 1 through 1 million.  You currently do not know whether the urn is small or large, but would like to find out.  Suppose you randomly draw one ball and find it numbered 7.  Given your draw, with what probability is the urn small?  

Let $T_1$ and $T_2$ represent the theories that the urn is small and large respectively, and let $D$ represent your data, i.e., that you have drawn the number 7.  Assuming you believe the urn equally likely to be small or large before you draw a number, then according to Bayes' Law, your updated belief in $T_1$ conditioned on observing your data is,
\begin{eqnarray*}
P(T_1|D) & = & \frac{P(D|T_1)P(T_1)}{P(D|T_1)P(T_1)+P(D|T_2)P(T_2)},\\
 & = & \frac{1/10 \times 1/2}{1/10 \times 1/2 + 1/10^6 \times 1/2},\\
 & \approx & 0.99999.
\end{eqnarray*}
You therefore should believe the urn is small with almost certainty.  Notice that you have become confident in the size of the urn with only one draw.

Now, multiply the number of balls in each urn by $10^{10}$ and imagine they correspond to birth numbers within our civilization, i.e., Adam drew number 1, Eve drew number 2, and so on (you and I have drawn numbers around $7\times10^{10}$).  $T_1$ and $T_2$ represent two competing theories, which we initially treat as equally likely.  Under $T_1$, the urn is small and contains only $10^{11}$ numbers, which means only $10^{11}$ people will ever exist and our civilization will die out within the next few centuries.  Under $T_2$, the urn is large and contains $10^{16}$ numbers so that our civilization is large and will continue on for many years into the future.  As before, you would like to determine whether the urn is small or large, i.e., whether our civilization is small or large. 

Given your data, $D=7\times10^{10}$, how should you update your belief in $T_1$?  Without any compelling argument to the contrary, you should update it as before, 
\begin{eqnarray*}
P(T_1|D) & = & \frac{P(D|T_1)P(T_1)}{P(D|T_1)P(T_1)+P(D|T_2)P(T_2)},\\
 & = & \frac{1/10^{11} \times 1/2}{1/10^{11} \times 1/2 + 1/10^{16} \times 1/2},\\
 & \approx & 0.99999,
\end{eqnarray*}
which means you should believe with almost certainty that our civilization is small and will die out within the next few centuries.

This argument is the Doomsday Argument (DA) as presented in Leslie (1989) and Leslie (1996).\footnote{Precursers of the argument are attributed to Carter (see Leslie (1996)) and Neilsen (1989); see \'{C}irkovi\'{c} (2004) for this specific treatment and Gott (1993) for an alternative description of the argument.}  The details of the DA can be restructured -- the numbers can be changed, more urn types can be added, etc.~-- but the final result remains unchanged:  when we condition on our birth number, we must drastically increase the probability that our civilization will soon die out.  

There are many critiques of the DA, which I will not focus on here (see Leslie (1996) or Bostrom (2002) for a full treatment).  By most accounts, the DA has stood up to all criticisms except one.  As first mentioned in Dieks (1992) and expanded in Bartha and Hitchcock (1999) and Olum (2002), the DA fails to consider that you are more likely to exist in a large civilization than a small one.  This missing step exactly cancels the updating of your beliefs so that your original prior is retained.

In this paper, I intend two things: (1) to defend the counter-argument to the DA developed in Bartha and Hitchcock (1999) and Olum (2002), and (2) to show that this counter-argument does not work when the DA is modified to allow for many worlds.  

The take-home message is the following: given that we exist in a civilization that has produced 70 billion people so far, we should drastically increase our belief that many other civilizations exist and that nearly all of these civilizations (including our own) will die out before producing trillions of people.

\section{The Devil's Existence}

\emph{The Devil's Existence} (DE) is a thought experiment where you are asked to determine whether or not the Devil exists (the Devil representing some doom event).  Suppose God creates 1 million rooms, each sequentially numbered, and that He initially intends to fill each room with one person.  He first goes to room 1 and generates a person inside.  He then moves on to room 2 and does the same, and so on, until the first 10 rooms have been filled.  At this point, if and only if the Devil exists, he arrives on the scene and destroys the remaining rooms so that no more people are created.  Suppose you have been created and find yourself in room 7.  Assuming you thought it equally likely that the Devil does or does not exist before considering this information, how likely is it now that the Devil exists?

In order to correctly update your belief in the Devil's existence, you must condition on all available information.  As stated above, your information consists of two things: (1) you exist (let $E$ represent that you exist), and (2) you are in room 7, (let $D$ represent that you are in room 7).  If $T_1$ and $T_2$ are the theories that the Devil does and does not exist, then according to Bayes' Law,
\begin{equation*}
P(T_1|D,E) = \frac{P(D|E,T_1)P(E|T_1)P(T_1)}{P(D|E,T_1)P(E|T_1)P(T_1)+P(D|E,T_2)P(E|T_2)P(T_2)}.
\end{equation*}
This equation requires that you know the probability of your existence given that the Devil does and does not exist, $P(E|T_1)$ and $P(E|T_2)$.  How do you calculate these?  One simple solution is to assume you are equally likely to exist independent of the Devil's existence, i.e., $P(E|T_1)=P(E|T_2)$.  When substituting $P(E|T_1)$ for $P(E|T_2)$ in the above equation, the probability of your existence drops out,
\begin{equation*}
P(T_1|D,E) = \frac{P(D|E,T_1)P(T_1)}{P(D|E,T_1)P(T_1)+P(D|E,T_2)P(T_2)}.
\end{equation*}
Now, your updated belief is easy to calculate.  Because you thought it equally likely that the Devil exists or does not exist, your priors are $P(T_1)=P(T_2)=1/2$.  Because you have a 1 in 10 chance of finding yourself in room 7 if there are 10 rooms, then $P(D|E,T_1)=1/10$.  Because you have a 1 in 1 million chance of finding yourself in room 7 if there are 1 million rooms, then $P(D|E,T_2)=1/10^6$.  Plugging these in,
\begin{eqnarray*}
P(T_1|D,E) & = & \frac{P(D|E,T_1)P(T_1)}{P(D|E,T_1)P(T_1)+P(D|E,T_2)P(T_2)},\\
& = & \frac{1/10 \times 1/2}{1/10 \times 1/2 + 1/10^6 \times 1/2},\\
& \approx & 0.99999.
\end{eqnarray*}
Therefore, after discovering that you are in room 7, you should believe with almost certainty that the Devil exists and your civilization is small.  Here, just as in the case of the urns, doom becomes very likely.

But, is it okay to assume you are equally likely to exist regardless of the Devil's existence?  Shouldn't you be more likely to exist if there are 1 million people created rather than only $10$?  I certainly have a better chance of winning the lottery if I play 1 million times rather than 10 times, and my existence seems much like the lottery.  Indeed, a large number of things could have been different such that my parents gave birth to someone other than me.  Out of the set of all the children they could have produced, I won.

Applying this thought process to the DE, suppose there exists a very large pool of possible people, $N$ in total, outside of the experiment.  When God generates a person, this means He randomly selects one person from the pool, embodies him/her, and then places the body in a room.  Bartha and Hitchcock (1999) and Olum (2002) argue that we should reason as if something like this was the case when analyzing the DE,\footnote{Olum argues this point using a thought experiment called \emph{God's Coin Toss}; Bartha and Hitchcock present a `just-so-story'.  Here, I have formulated DE because it allows me to correctly analyze the possibility of many worlds later in the paper.} and I agree.  I adopt their stance and leave a defense of this stance until the next section.  

Considering existing people as random pulls from the set of $N$ possible people, the probabilities that you are created in the case of the Devil existing and not existing are,\footnote{I assume these pulls are with replacement.  If $x$ is the number of existing people in theory $T$, then $P(E|T)=1-(1-1/N)^x$.  When $N\gg x$, as assumed here, then $P(E|T)\approx x/N$.  When the number of existing people is much larger than the number of possible people, $x \gg N$, as will happen when considering many worlds, then $P(E|T)\approx 1$.} 
\begin{equation*}
P(E|T_1) \approx 10/N,
\end{equation*}
and
\begin{equation*}
P(E|T_2) \approx\ 10^6/N.
\end{equation*}
You have approximately a 10 in $N$ chance of existing if 10 people are created in total and a 1 million in $N$ chance of existing if 1 million people are created.  Plugging these into Bayes' Law,
\begin{eqnarray*}
P(T_1|D,E) & = & \frac{P(D|E,T_1)P(E|T_1)P(T_1)}{P(D|E,T_1)P(E|T_1)P(T_1)+P(D|E,T_2)P(E|T_2)P(T_2)},\\
& \approx & \frac{1/10 \times 10/N \times 1/2}{1/10 \times 10/N \times 1/2 + 1/10^6 \times 10^6/N \times 1/2},\\
& \approx & 1/2.
\end{eqnarray*}
Therefore, discovering that you exist and are in room 7 does not cause a shift in your beliefs, and you remain undecided about the Devil's existence.  The Doomsday Argument has failed and doom is averted.  

Why did the Doomsday Argument fail?  You are much more likely to discover your room number is 7 under $T_1$ than $T_2$: there is a 1 in 10 chance of being in room 7 if the Devil exists and a 1 in 1 million chance if the Devil does not exist.  Therefore, finding that your room number is 7 makes the Devil's existence 100,000 times more likely than his non-existence.  But, you should also take into account that you are more likely to exist under $T_2$ than $T_1$: 1 million people are created if the Devil does not exist and only 10 people are created if the Devil does exist.  Therefore, knowing that you exist makes it 100,000 times more likely that the Devil does not exist.  These two probability shifts -- one due to finding yourself in room 7 and the other due to your existence -- cancel each other so that you retain your original beliefs. 

If at this point you feel confident that the Doomsday Argument does not work, you should not remain comfortable for long.  If you believe it possible that many civilizations exist (perhaps in other worlds), then doom will return.  

\section{The Presumptuous Philosopher}

Although several arguments against Bartha and Hitchcock (1999) and Olum (2002) exist (see Bostrom and \'{C}irkovi\'{c} (2003) and \'{C}irkovi\'{c} (2004)), there is one argument that has received the most attention.  Consider \emph{The Presumptous Philosopher} (PP) as described in Bostrom (2002):
\begin{quote}
It is the year 2100 and physicists have narrowed down the search for a theory of everything to only two remaining plausible candidate theories, $T_1$ and $T_2$ (using considerations from super-duper symmetry). According to $T_1$ the world is very, very big but finite and there are a total of a trillion trillion observers in the cosmos. According to $T_2$, the world is very, very, very big but finite and there are a trillion trillion trillion observers. The super-duper symmetry considerations are indifferent between these two theories. Physicists are preparing a simple experiment that will falsify one of the theories. Enter the presumptuous philosopher: ``Hey guys, it is completely unnecessary for you to do the experiment, because I can already show to you that $T_2$ is about a trillion times more likely to be true than $T_1$!''
\end{quote}
The point is the following.  In the DE thought experiment, I stated we should reason as though each existing person is a random pull from the set of possible people.  Bostrom believes this reasoning is wrong because it forces anyone adopting it to prefer certain theories over others simply because they produce more people.  He reckons that if $T_1$ and $T_2$ are otherwise just as likely, but $T_2$ contains 1 trillion times as many people as $T_1$, then I must believe that $T_2$ is 1 trillion times more likely than $T_1$ simply due to this fact (which leads to presumptuous conclusions).  Unfortunately, Olum agrees with Bostrom on this point,
\begin{quote}
For example, suppose I have a crazy theory that each planet actually has $10^{{10}^{100}}$ copies of itself on 'other planes'. Suppose that (as cranks often do) I believe this theory in spite of the fact that every reputable scientist thinks it is garbage. I could argue that my theory is very likely to be correct, because the chance that every reputable scientist is independently wrong is clearly more than 1 in $10^{{10}^{100}}$.  To avoid this conclusion, one must say that the a priori chance that my theory is right is less than 1 in $10^{{10}^{100}}$. It seems hard to have such fantastic confidence that a theory is wrong, but if we do not allow that, we shall be prey to the argument above.
\end{quote}

But, must we always prefer theories that contain more people if we reason as though existing people are random pulls from the set of possible people?  No.  Here is why:

\subsection{Sometimes it is Bostrom who is presumptuous}

We must be very careful when analyzing theories that posit the existence of unknown people.  For example, suppose the extra people that exist in $T_2$ possess some characteristic, $C$, that you, I, and the other people in $T_1$ do not possess.  Then $T_2$ would not be more likely than $T_1$.  The increased probability of existence in $T_2$ is exactly cancelled by the decreased probability of not possessing $C$ in $T_2$.  In the DE thought experiment above, $C$ is `having a birth number greater than 10'.  As shown above, you do not prefer $T_2$ to $T_1$ in the DE even though $T_2$ produces more people.

As another example, suppose you are contemplating whether jovians exist, i.e., whether there are inhabitants of Jupiter.\footnote{This example is first discussed in Neal (2006) and Hartle and Srednicki (2007).}  If they do exist, then there are more ``people'', so you should increase your belief in their existence (assuming jovians are as equally likely as humans to be you).   However, jovians all share the same characteristic $C=$ `being jovian' that you and I do not possess.  These effects cancel one another so that you retain your original prior (see Garriga and Vilenkin (2008)).  A philosopher who agrees with Bostrom, however, would not retain her original prior.  Consider \emph{The Presumptuous Philosopher II}:

\begin{quote}
Suppose that in the not too distant future we discover evidence of a vast alien civilization that once ruled our galaxy but that is now extinct.  After considering all of the evidence, scientists believe that only two candidate theories about the civilization are plausible, $T_1$ and $T_2$.  According to $T_1$, the civilization was very, very large and produced a total of a trillion trillion observers.  According to $T_2$, the civilization was very, very, very large and produced a total of a trillion trillion trillion observers.  Scientists currently believe both these theories are equally likely and are preparing a simple experiment that will falsify one of them.  Enter the presumptuous philosopher: ``Hey guys, it is completely unnecessary for you to do the experiment, because I can already show to you that $T_1$ is about a trillion times more likely to be true than $T_2$!''
\end{quote}

For the philosopher who agrees with Bostrom, if $T_1$ is true, it does not make her any more likely to exist.  However, it does make her approximately 1 trillion times more likely to be one of the $7\times10^{10}$ humans who have existed (rather than one of the aliens).  Because, of course, she is human, the philosopher believes $T_1$ to be about a trillion times more likely than $T_2$ -- a presumptuous belief that Bostrom shares with the philosopher but that the original presumptuous philosopher and I do not.

\subsection{Presumptuous assumptions}

\emph{The Presumptuous Philosopher} (PP) depends on two assumptions which are both rather presumptuous: (1) the PP assumes the philosopher is certain her existence is highly atypical under $T_1$ and (2) the PP assumes the philosopher is certain all alternative theories which produce many people are false.

For the PP to work, the philosopher must be certain that her existence is atypical under $T_1$, i.e., she must certain that the number of possible people is much larger than a trillion trillion, $N \gg 1$ trillion trillion.  Suppose that $N < 1$ trillion trillion so that the philosopher is likely to exist under both theories; then she would not strongly prefer $T_2$ to $T_1$.  $P(E|T_1)\approx P(E|T_2) \approx 1$, and she approximately retains her prior belief in both theories, $P(T_1|E)\approx P(T_1)$ and $P(T_2|E) \approx P(T_2)$.\footnote{Neal (2006) was the first to point out that if you believe the universe is large enough to make your existence likely, then you neither favor nor disfavor theories that posit an even larger universe.}   

To make this point more explicit, suppose that -- surprise, surprise -- physicists ignore the presumptuous philosopher and perform their experiment anyway.  They find, to the philosopher's astonishment, that $T_2$ is false and the world is not very, very, very big.  Would the philosopher then think, ``The experiment was 1 trillion times more likely to find $T_1$ false than $T_2$.  We just have to accept that our universe is very special, 1 trillion times more special than it would have been had we found $T_1$ false.''  Bostrom thinks so, but I do not.  Instead, she is much more likely to question her original assumptions such as her belief that the number of possible people is larger than a trillion trillion.  Keep in mind that a possible person is not the same as possible instances of a possible person.  I could have had red hair instead of brown, or perhaps even been a jovian instead of a human, etc.  In these alternative cases, of which there are many, I am different instances of the same possible person and account for only 1 out of the $N$ number of possible people.  Although it may seem unlikely, perhaps $N$ really is less than a trillion trillion.

Let us suppose that $N$ is certainly larger than a trillion trillion so that the previous considerations are not important.  The PP still requires that the philosopher be certain \emph{all} alternative theories which produce many people are false.  As before, suppose that physicists have shown $T_2$ false -- a result they repeat over and over so that we are certain it is correct.  Must the philosopher then reluctantly accept $T_1$ true, all the while believing $T_1$ a 1 in 1 trillion fluke?  Of course not.  She is much more likely to accept alternatives to $T_1$ that produce many people, such as the many worlds interpretation of quantum mechanics or the existence of multiple universes.  For example, suppose the philosopher believes it plausible that the cosmos is exactly as $T_1$ describes, but also with many quantum worlds.  Then the philosopher would not be so surprised to find that $T_2$ is false.  There are plenty of people in ``$T_1 + $many worlds'' that her existence is already likely.  For the PP argument to work, the philosopher must be certain (up to 1 trillion to 1 odds) that quantum many worlds (and \emph{any other} theory that produces many people) is incorrect.  It is hard to justify such certainty that these theories are false.

The presumptuous philosopher, therefore, is presumptuous not because she reasons as though existing people are pulls from the set of possible people, instead she is presumptuous because she is certain that the number of possible people is larger than a trillion trillion and because she is certain that every other theory that generates many people is false.

\subsection{It is already plausible that you exist}

If you believe, as many cosmologists do, that there is not one single universe, but many universes collectively referred to as the multiverse and furthermore, that there are many quantum worlds branching out of each of these universes, then it is almost certain the number of realized people is larger than $N$ and there are multiple instances of you (Garriga and Vilenkin (2008), Page (2010), and Srednicki and Hartle (2010)).  

Given your existence is already plausible, ``crazy'' ideas such as Olum's $10^{10^{100}}$ planet-multiplier theory are not needed to make your existence more likely.  It is \emph{not} true your prior belief in Olum's theory needs to be less than $1/10^{10^{100}}$ in order to dismiss it.  Priors as low as $1/10^3$ will do, as long as you agree your existence is already likely without use of Olum's theory.

\section{The Devil's Existence in Many Worlds}

Suppose \emph{The Devil's Existence} (DE) thought experiment is updated to allow for the possibility that many civilizations exist.  There is either a single civilization or many civilizations.  If there is a single civilization, then the experiment is run once exactly as before.  If there are many civilizations, then the experiment is run simultaneously in many parallel worlds with many civilizations created -- so many in fact, that every possible person is almost sure to exist.  If the Devil exists (i.e., if civilizations tend to be small), then he destroys the excess rooms in each of the worlds so that all civilizations contain 10 people.  If the Devil does not exist (i.e., if civilizations tend to be large), then all rooms are filled and all civilizations contain 1 million people.

There are four possibilities.  Either the Devil exists and there is a single civilization (label the theory that this is true $T_{1,S}$), the Devil exists and there are many civilizations ($T_{1,M}$), the Devil does not exist and there is a single civilization ($T_{2,S}$), or the Devil does not exist and there are many civilizations ($T_{2,M}$).   Your task is to determine which of these four theories is correct.  Suppose, as before, you exist and find yourself in room 7.  Assuming you initially believe the theories equally likely before considering this information, which of the four theories is now most likely to be true?

Because many, many people are created in $T_{1,M}$ and $T_{2,M}$, it is almost certain that you exist if one of these theories is correct.  Furthermore, it is likely there are many instances of you.  In the calculations that follow, I assume that if you exist, your current instance, i.e., the existing person in which you are currently embodied, is chosen randomly from the set of all of your instances.  Notice that this assumption is different from (and in addition to) reasoning as though existing people are random pulls from the set of possible people.\footnote{This distinction is very important.  Suppose that if the Devil does not exist, then God generates monkeys in the last 999,990 rooms (see Olum (2002)).  I can still assume that my current instance is chosen randomly from the set of all of my instances, but I could (and probably should) assign a very low probability that a monkey is one of my instances.  When you split your reasoning into two parts: (1) determining when and where your instances exist and (2) determining which instance of ``you'' you currently experience, then the reference class problems discussed in Bostrom (2002) disappear.  Instead, you are left with the much more palatable problem of assigning probabilities that a particular type of object is one of your instances.}

To update your beliefs, you need an estimate for $N$.  Here, I will assume that $N=2^{10^{15}}$, which is based on the number of possible neural connections in the human brain (see Page (2010)).  The actual number is not of consequence below, only the assumption that it is a very large but finite number is important.  

Define the following four probabilities,
\begin{eqnarray*}
P_1 & \equiv & P(D|E,T_{1,S})P(E|T_{1,S})P(T_{1,S}),\\
 & \approx & 1/10 \times 10/2^{10^{15}} \times 1/4,\\
 & \approx & 0.
\end{eqnarray*}
\begin{eqnarray*}
P_2 & \equiv & P(D|E,T_{1,M})P(E|T_{1,M})P(T_{1,M}),\\
 & \approx & 1/10 \times 1 \times 1/4,\\
 & \approx & 1/40.
\end{eqnarray*}
\begin{eqnarray*}
P_3 & \equiv & P(D|E,T_{2,S})P(E|T_{2,S})P(T_{2,S}),\\
 & \approx & 1/10^6 \times 10^6/2^{10^{15}} \times 1/4,\\
 & \approx & 0.
\end{eqnarray*}
\begin{eqnarray*}
P_4 & \equiv & P(D|E,T_{2,M})P(E|T_{2,M})P(T_{2,M}),\\
 & \approx & 1/10^6 \times 1 \times 1/4,\\
 & \approx & 1/(4\times10^6).\\
\end{eqnarray*}
According to Bayes' Law,
\begin{equation*}
P(T_{1,S}|D,E) =  \frac{P_1}{P_1+P_2+P_3+P_4} \approx 0,
\end{equation*}
\begin{equation*}
P(T_{1,M}|D,E) =  \frac{P_2}{P_1+P_2+P_3+P_4} \approx 0.99999,
\end{equation*}
\begin{equation*}
P(T_{2,S}|D,E) =  \frac{P_3}{P_1+P_2+P_3+P_4} \approx 0,
\end{equation*}
\begin{equation*}
P(T_{2,M}|D,E) =  \frac{P_4}{P_1+P_2+P_3+P_4} \approx 0.00001.
\end{equation*}
Therefore, after discovering that you exist and are in room 7, you should believe with almost certainty that there are many civilizations and that the Devil exists so that all of these civilizations are small.

The above argument is immune to the criticism that Bartha and Hitchcock (1999) and Olum (2002) level against the original Doomsday Argument.\footnote{However, note that if we assume the total number of existing people, $x$, is \emph{always} much less than the number of possible people, $N$, no matter how many civilizations exist, then the Doomsday Argument (even in many worlds) would fail.}  In fact, the above argument depends critically on their insight -- that when analyzing the DA, we should reason as though existing people are random pulls from the set of possible people. 

\subsection{Are many civilizations plausible?}

How plausible is it that many civilizations exist?  In the above argument I assumed your priors were all equal, meaning you initially believed it equally likely that there are many, many civilizations or only one.  Although we have no evidence that other civilizations exist, it is still very plausible that they do.  From what we know of our vast universe, there appears to be sufficiently many other worlds (causally separated regions of space-time) such that many, many other civilizations exist (Garriga and Vilenkin (2008), Page (2010), and Srednicki and Hartle (2010)).

Even if you initially think it implausible that many civilizations exist, then as long as you agree to the following:
\begin{itemize}
\item[(1)] There are a very, very large but finite number of possible people.\footnote{For example, you might believe that a particular complex arrangement of matter specifies each possible person and that there are many such arrangements corresponding to many possible people.  Presumably for humans, the complex arrangement of matter that determines who is in which body is located somewhere in the brain.}
\item[(2)] You are an instance of one of these possible people.
\item[(3)] You exist in a civilization that has so far produced $7\times10^{10}$ people.
\end{itemize}
then reasoning as though existing people are random pulls from the set of possible people forces you to drastically increase your initial belief (whatever it is) that many other civilizations exist.  This does not mean you are forced to accept any ``crazy'' theory that produces more civilizations with more people.  Instead it means you should increase your belief in the general theory that there are many other civilizations.  Exactly how and where these civilizations exist is up for debate.

\subsection{Alternative (incorrect) treatments of the DA in many worlds}

Both Leslie (1996) and Olum (2002) briefly consider a many worlds version of the Doomsday Argument.  In the context of the DE, their analysis amounts to assuming the following:  God fills the first 10 rooms with people and then flips a fair coin to determine whether or not the Devil exists (heads, he exists and tails, he does not).  As before, if the Devil exists, he destroys the remaining rooms.  Many worlds is interpreted as running the experiment many times, each time with a separate coin flip.  Your task is to determine whether the coin landed heads or tails in your run of the experiment.

Unfortunately, the setup that Leslie and Olum adopt is not very useful.  It \emph{assumes} the cosmos produces civilizations that have a 1/2 chance of experiencing a doom event.  Suppose the cosmos is different such that doom events are nearly certain and most civilizations are small, or alternatively, that doom events are rare so that most civilizations are large, or possibly even that doom events have a 3/4 chance of occurring.  Unfortunately, you cannot consider any of these possibilities.  You are forced to accept that doom for your civilization is determined by the flip of a fair coin.  Faced with this imposed reality, your task is to discern whether the coin toss lands heads or tails for your civilization.  But here, again, the experiment makes little sense.  It is rather silly to ask someone, ``With what probability will a fair coin toss land heads?'', yet we ask just this question to everyone in rooms 1 through 10.  Perhaps we should not be surprised when their answer is 1/2.

Instead of asking the participants the outcome of a fair coin toss, we should be asking them their belief in different theories about the bias of the coin.  When formulating the DE in many worlds, I considered only two possibilities for the existence of a doom event.  Either the probability of doom is zero and all civilizations are large or the probability of doom is one and all civilizations are small.  Of course, if I wanted to be thorough, I would have included many other possibilities, including the possibility that doom occurs with probability 1/2 (which Leslie and Olum assume).  Suppose I added other possibilities for you to decide between.  The final result would remain unchanged: after considering your circumstance (that you exist and are in room 7), you would drastically increase your belief that many civilizations exist and nearly all of these civilizations are small.\footnote{Suppose I included $T_{3,S}$ and $T_{3,M}$, which both state that a doom event occurs with probability 1/2 on each run of the experiment, but with a single and many civilization(s) respectively.  If all six theories are initially equally likely, then $P(T_{3,S}|D,E)\approx0$, $P(T_{3,M}|D,E)\approx0.00002$, and the clear choice would still be $T_{1,M}$ with $P(T_{1,M}|D,E)\approx0.99997$.}

\subsection{Universal Doomsday}

Knobe, Olum, and Vilenkin (2006) present a \emph{Universal Doomsday Argument} where they assume the universe is infinite (with many civilizations) and then show that given our birth number, we should drastically increase our belief that almost all civilizations are short-lived.  Their conclusion is nearly identical to what I state here, but I have reservations about their analysis.  They use Bayes' Law to update the estimate of the frequency of short-lived civilizations in our universe, i.e., the probability that a randomly selected civilization in our universe is short-lived.  Let $P(S)$ be our estimate of this probability and let $P(L)=1-P(S)$ be our estimate of the probability that a randomly selected civilization in our universe is long-lived.  Let $x_S$ be the number of people in a short-lived civilization, let $x_L$ be the number of people in a long-lived civilization, and assume $x_L\gg x_S$.  Our data, $D$, is our birth number, $7\times10^{10}$.  They state,
\begin{equation*}
P(S|D) = \frac{P(D|S)P(S)}{P(D|S)P(S)+P(D|L)P(L)}.
\end{equation*}
and furthermore that,
\begin{eqnarray*}
P(D|S) & = & 1/x_S,\\
P(D|L) & = & 1/x_L.
\end{eqnarray*}
Since $x_L\gg x_S$, $P(D|S)\gg P(D|L)$ and thus $P(S|D)$ is nearly 1 unless $P(S)$ is extremely small.  Therefore, according to Knobe, Olum, and Vilenkin, we should believe that nearly all civilizations are small when conditioning on our birth number.  However, they have used $S$ and $L$ to denote different things at different points in their analysis.  In their priors, $S$ and $L$ denote that a randomly selected civilization in our universe is short-lived and long-lived respectively.  In their likelihoods, $S$ and $L$ denote that \emph{our} civilization is short-lived and long-lived respectively.  If we make $S$ and $L$ consistent across the priors and likelihoods, then the final result is ruined.

Instead of using Bayes' Law to update our estimate of the frequency of short-lived and long-lived civilizations in our universe (as Knobe, Olum, and Vilenkin attempt), I think we should update our beliefs in different theories about the frequency of short-lived and long-lived civilizations in our universe.  For example, suppose I believe that either all civilizations in our universe are short-lived, $T_S$, or that all civilizations in our universe are long-lived, $T_L$.  I don't know which of these theories is correct and would like to update my prior belief in both theories after considering my birth number.  Suppose my prior belief in $T_S$ is $P(T_S)$, and my prior belief in $T_L$ is $P(T_L)=1-P(T_S)$.  Therefore,
\begin{equation*}
P(T_S|D) = \frac{P(D|T_S)P(T_S)}{P(D|T_S)P(T_S)+P(D|T_L)P(T_L)}.
\end{equation*}
and 
\begin{eqnarray*}
P(D|T_S) & = & 1/x_S,\\
P(D|T_L) & = & 1/x_L.
\end{eqnarray*}
Since $x_L\gg x_S$, $P(D|T_S)\gg P(D|T_L)$ and thus $P(T_S|D)$ is nearly 1 unless $P(T_S)$ is extremely small.  If Knobe, Olum, and Vilenkin had presented their argument in this way, then I would agree with their statement, ``we should now think that almost all civilizations will be short-lived -- a sort of `universal doomsday'.''

\section{Conclusion}

Consider our current circumstance: we exist in a civilization that has produced only 70 billion people.  At first sight, our data seems highly unlikely.  If our civilization dies out soon so that it is ultimately small, then not that many people exist and it is highly unlikely that you and I are alive.  If our civilization is ultimately very, very large, then our existence might be explained, but it is highly unlikely for you and I to have such low birth numbers.  Given the situation, we can either accept that we are atypical or we can seek plausible alternative theories that better explain our data.

Here, I have considered the following theory: there are many, many civilizations that exist and nearly all of these civilizations are small.  Under this theory, our existence is certain and our birth number is typical.  Because the theory makes our circumstance likely and alternatives do not, we should drastically increase our belief that the theory is true when conditioning on our data.

\section*{Acknowledgements}

I thank Paul Bartha, Milan \'{C}irkovi\'{c}, and Ken Olum for very helpful comments and suggestions.


\begin{thebibliography}{16}

\bibitem {B0} Bartha, P. and Hitchcock, C. (1999) ``No One Knows the Date or the Hour: an Unorthodox Application of Rev. Bayes' Theorem'', \emph{Philosophy of Science (Proceedings)}, {\bf 66}, S229.

\bibitem {B1} Bostrom, N. (2002) \emph{Anthropic Bias: Observation Selection Effects}, Routledge, New York.

\bibitem {B2} Bostrom, N. and \'{C}irkovi\'{c}, M. M. (2003) ``The Doomsday Argument and the Self-Indication Assumption: Reply to Olum'', \emph{Philosophical Quarterly}, {\bf 53}, 83.

\bibitem {C1} \'{C}irkovi\'{c}, M. M. (2004) ``Is Many Likelier than Few? A Critical Assessment of the Self-Indication Assumption'', \emph{Epistemologia}, {\bf 27}, 265.

\bibitem {D1} Dieks, D. (1992) ``Doomsday -- Or: The Dangers of Statistics'', \emph{Philosophical Quarterly}, {\bf 42}, 78.

\bibitem {G1} Garriga, J. and Vilenkin, A. (2008) ``Prediction and Explanation in the Multiverse'', \emph{Physical Review D}, {\bf 77}, 043526.

\bibitem {G2} Gott, J. R. (1993) ``Implications of the Copernican Principle for our Future Prospects'', \emph{Nature}, {\bf 363}, 315.

\bibitem {H1} Hartle, J. B. and Srednicki, M. (2007) ``Are We Typical?'' \emph{Physical Review D}, {\bf 75}, 123523.

\bibitem {K1} Knobe, J., Olum, K. D., and Vilenkin, A. (2006) ``Philosophical Implications of Inflationary Cosmology'', \emph{British Journal for the Philosophy of Science}, {\bf 57}, 47.

\bibitem {L1} Leslie, J. (1989) ``Risking the World's End'', \textit{Bulletin of the Canadian Nuclear Society}, May 1989, 10.

\bibitem {L2} Leslie, J. (1996) \emph{The End of the World: The Ethics and Science of Human Extinction}, Routledge, London.

\bibitem {N1} Neal, R. M. (2006) ``Puzzles of Anthropic Reasoning Resolved Using Full Non-Indexical Conditioning'', arXiv:math/0608592

\bibitem {N2} Neilsen, H. B. (1989) ``Random Dynamics and Relations between the Number of Fermion Generations and the Fine Structure Constants'', \emph{Acta Physica Polonica B}, {\bf 20}, 427.

\bibitem {O1} Olum, K. D. (2002)  ``The Doomsday Argument and the Number of Possible Observers'', \textit{Philosophical Quarterly}, {\bf 52}:207, 164.

\bibitem {P1} Page, D. N. (2010) ``Our Place in a Vast Universe'', In \emph{Science and Religion in Dialogue} (ed M. Y. Stewart), Wiley-Blackwell, Oxford.

\bibitem {S1} Srednicki, M. and Hartle, J. (2010) ``Science in a Very Large Universe'', \emph{Physical Review D}, {\bf 81}, 123524.

\end{thebibliography}
\end{document}